 \documentclass[apj]{emulateapj}

\usepackage{natbib}
\usepackage{graphicx}
\bibpunct{(}{)}{;}{a}{}{,}

\begin{document}

   \title{Magnetic bright points in the quiet Sun}

   \author{
   	J.~S\'anchez~Almeida\altaffilmark{1,2},
 	J.~A.~Bonet\altaffilmark{1,2},
 	B.~Viticchi\'e\altaffilmark{3,4},
 	D.~Del~Moro\altaffilmark{4}
 	}
    \altaffiltext{1}{Instituto de Astrof\'\i sica de Canarias, 
              E-38205 La Laguna, Tenerife, Spain}
    \altaffiltext{2}{Departamento de Astrof\'\i sica, Universidad de La Laguna,
        E- 38071 La Laguna, Tenerife, Spain}
    \altaffiltext{3}{ESA/ESTEC RSSD, Keplerlaan 1, 2200 AG Noordwijk, 
        Netherlands}
    \altaffiltext{4}{Dipartimento di Fisica, Universit\`a degli Studi di Roma "Tor Vergata",
        I-00133 Rome, Italy}
   \email{jos@iac.es, jab@iac.es, Bartolomeo.Viticchie@esa.int,
        delmoro@roma2.infn.it}

\begin{abstract}
We present a visual determination of the number  
of bright points (BPs) existing in the quiet Sun, which 
are structures though to trace intense kG magnetic concentrations. 
The measurement is based on a  0\farcs 1 angular resolution G-band 
movie obtained  with the Swedish Solar Telescope at the solar disk center. 
We find 0.97 BPs~Mm$^{-2}$, which
is a factor three larger than any previous estimate.
It corresponds to  1.2 BPs per solar granule.
Depending on the details of the segmentation, the BPs
cover between 0.9\,\% and 2.2\,\% of the solar surface. 
Assuming their field strength to be 1.5 kG, 
the detected BPs contribute to the solar magnetic flux with 
an unsigned flux density between 13~G and 33~G.
If network and inter-network regions are counted separately,
they contain 2.2 BPs~Mm$^{-2}$ and 0.85  BPs~Mm$^{-2}$, 
respectively.
\end{abstract}



\shorttitle{Magnetic Bright Points in the quiet Sun}
\shortauthors{S\'anchez Almeida  et al.}

\section{Motivation}\label{intro}

Our understanding of the quiet Sun magnetic magnetic fields
has turned over during the last decade. 
Traditional magnetograms showed magnetic signals
only at the network boundaries, occupying only a small
fraction of the solar surface, and mainly produced
by spatially unresolved kG magnetic concentrations 
\citep[e.g.,][]{bec77b,sol93}.
Nowadays magnetic signals are detected almost 
everywhere 
\citep[e.g.,][]{dom03a,har07,lit08}, 
and there is no doubt on the presence of a volume filling 
component of dG and hG fields as inferred from the Hanle 
depolarization signals \citep[e.g.,][]{fau93,tru04}. 
The intra-network (IN) magnetic fields have passed
from hardly delectable signals in long-integration 
low spatial resolution magnetograms 
\citep[e.g.,][]{mar88,wan95},
to ubiquitous features \citep[e.g.,][]{lin99,san00}.
The abundance of new magnetic structures makes them  potentially important to 
understand the global magnetic properties of the Sun 
\citep[][]{san04,tru04}, 
and also makes it unlikely that the quiet Sun magnetism results 
from the decay of active regions \citep[e.g.,][]{san03b}.
Since the quiet Sun magnetic fields do not seem to be
debris from active regions, they are not automatically 
generated by the dynamo responsible for the solar cycle. 
A different production mechanism seems 
to be at work, probably
an efficient turbulent 
dynamo driven by granular convection 
\citep{pet93,cat99b}, an hypothesis corroborated by the 
latest numerical simulations of granular magneto-convection 
\citep[][]{vog07,pie10}.
The sensitivity of the current polarimeters allow us
to detect weak linear polarization signals 
produced by the transverse component of the 
magnetic fields \citep[][]{har07,lit08}, 
and to find out that they often emerge as 
short loops pooping up from the sub-photosphere 
\citep[][]{cen07,mar09}.
In agreement with the turbulent dynamo scenario,
quiet Sun magnetic fields come with strengths 
in the full range from basically zero to 2\,kG 
\citep[][]{san00,dom06}. Even if they only
fill a small fraction of the quiet photosphere,
the part having strong kG fields may be particularly 
important from a physical point of view. 
Magnetic flux and magnetic energy scale as powers of 
the field strength, therefore, depending on 
their (uncertain) area coverage, kGs may surpass 
the contribution of  the more common but weaker fields
\citep[][]{san04}.
Moreover, buoyancy makes kG concentrations 
vertical \citep[e.g.,][]{sch86} and so, they can provide 
a mechanical connection between the photosphere and the upper
atmosphere \citep[e.g.,][]{bal98,san08}. 
Theoretical studies support the traditionally-ignored but 
potentially-important influence of the quiet Sun 
photosphere on the chromosphere, transition region, 
and corona \citep[e.g.][]{sch03b,goo04,jen06}.
Despite all these (and other) significant advances, 
our understanding of the quiet Sun magnetism is still 
incomplete.  
Each observational improvement reveals 
more magnetic structures, in a process showing no 
sign of having converged yet.

As a part of this movement to characterize the properties of the 
quiet Sun magnetism, we discovered the ubiquitous presence of 
bright points (BPs) in the intergranular lanes of the 
quiet Sun \citep{san04a}. The plasma in the lanes 
is cold and therefore dark, but the BPs look bright because they trace intense 
kG magnetic concentrations \citep[]{spr77}. The magnetic field provides 
most of the pressure required to maintain the structure in mechanical 
balance, which reduces the gas pressure, the density and, thus, 
the opacity. The kG concentrations resemble holes carved on the solar 
surface which allow us to peep up into the sub-photosphere, where  
deeper usually means hotter and brighter. This picture is confirmed by 
realistic numerical simulations \citep[e.g.,][]{car04,kel04}, 
and thus BPs are regarded as proxies for kG magnetic concentrations.
The presence of BPs everywhere on
the quiet Sun has been verified 
in several works, e.g., \citet[][]{dew05,dew08,bov08,san08}.
However, none of these works find as many BPs as 
\citet[][0.3 BPs\,Mm$^{-2}$]{san04a}. This is 
partly due to the angular resolution of the images, 
taken with small telescopes, and also due to the use 
of automatic algorithms for BP identification which can 
hardly detect the faintest BPs.
In an effort to calibrate  with visual detection
one of these methods \citep[][]{bov01,bov07}, 
we repeated the tedious but otherwise accurate analysis 
carried out by \citet{san04a}. This time we use a 
G-band image of much better quality. 
The result of such detailed analysis is surprising, 
and we report it here. We find three times more 
G-band BPs than in the original study and, therefore,
far more than the numbers reported in all the previous 
studies cited above.  Our density implies that the quiet Sun 
has more BPs than granules,
and at least 1\,\% of the quiet solar surface is 
covered by 
BPs.

\section{Observation and analysis procedure}\label{description}

\begin{figure*}
\includegraphics[width=0.9\textwidth,angle=90]{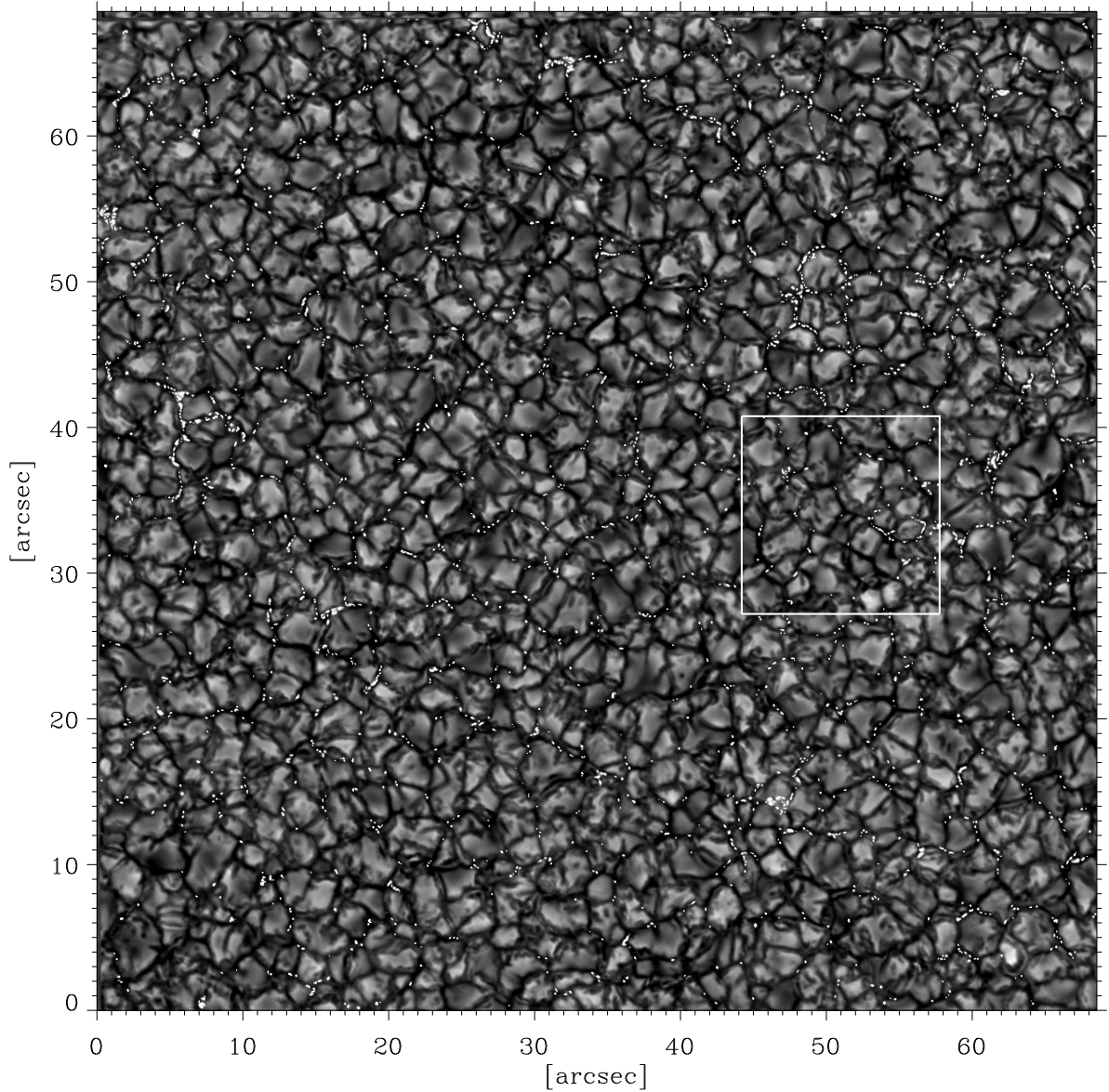}
\caption{
G-band image with selected BPs enhanced. 
BPs are found throughout, with a number density of 
0.97 BPs~Mm$^{-2}$ or, equivalently, of 1.2~ BPs~per~granule.
The square box outlines the region blown up in Figs.~\ref{zoom}. 
}
\label{fullfov}
\end{figure*}

Figure~\ref{fullfov} shows the image selected for in-depth 
analysis. It is the best in a series of quiet Sun disk 
center images taken with the Swedish Solar Telescope 
\citep[SST; ][]{sch02} in the G-band (a 10.8~\AA\ wide filter 
centered at 4305.6~\AA). The $68\farcs 5\times 68\farcs 5$
field of view (FOV) was observed on September 29, 2007. 
The series was restored 
using multi-frame blind deconvolution \citep[][]{noo05} 
to have an angular resolution close to the diffraction
limit of the SST at the working wavelength 
($\simeq$0\farcs 1).  Each snapshot results from combining 
125 images, rendering a mean cadence of one snapshot 
every 15~s. The dataset includes simultaneous Ca\,H line-core
images, which we use to determine network boundaries.
Further details are given by \citet[][]{bon08}.

The BPs were selected in a way similar to that described
by \citet{san04a}. The reference image was segmented into
locally bright features and background using the 
algorithm by \citet{str94}. We flick on the computer screen 
the segmented image and the true image, selecting as BPs 
bright structures coinciding with one of the segmented 
patches. 
\begin{figure}
\includegraphics[width=0.4\textwidth,angle=90]{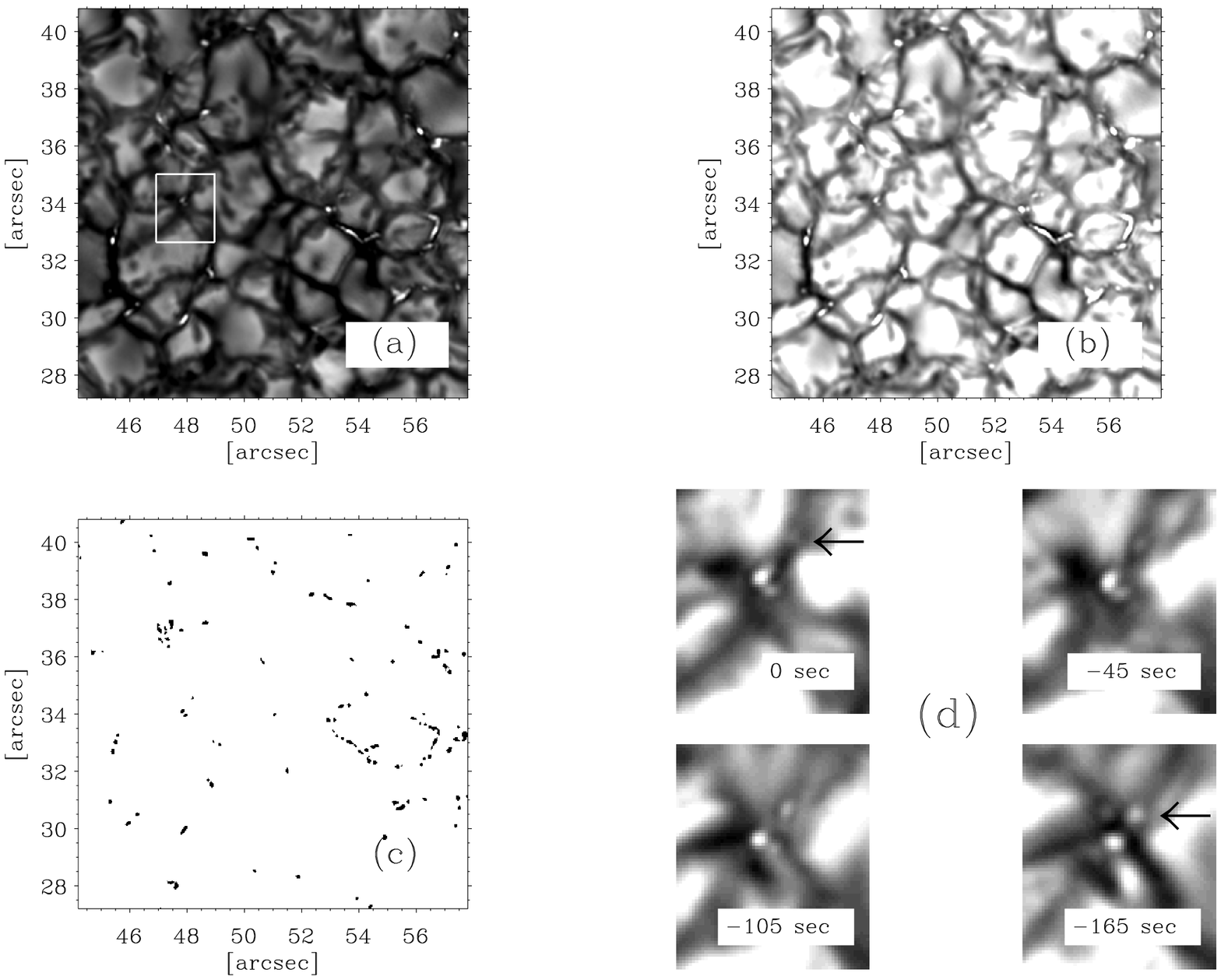}
\caption{Illustration of the visual selection
procedure. (a) Low contrast image of a small subfield 
of the FOV (see the box in Fig.~\ref{fullfov}). It is used 
to select the brightest G-band BPs. 
(b) Same image as (a) shown with high contrast.
Used to identify fainter BPs. (c) Binary map with the selected
structures. (d) Time series (see the insets) illustrating how
an uncertain BP in the reference image 
(time 0~sec, the arrow) was actually a clear and conspicuous 
BP a few minutes before (time -165~sec, the arrow). 
The subfield shown in the time series is marked as a box in (a).
}
\label{zoom}
\end{figure}
We first pinpoint obvious BPs in the intergranular 
lanes of the best image.  In case of doubt, we play back and 
forth the animation with snapshots around it.  We then select as BPs in our image 
those that along the time sequence show up as clean BPs (see the
example in Fig.~\ref{zoom}d). Doubtful identifications were 
discarded.  Moreover, we also include BPs that were conspicuous 
before or after, but  which remain as 
faint brightenings in the reference snapshot because 
of evolution.
The procedure was repeated using two different 
saturation levels for the true image. 
The low contrast (between 0.8 and 1.5, in units 
of the mean quiet Sun intensity) was used to select the 
brightest BPs, whereas the high contrast (between 0.7 and 1.1) 
was used for dimmer BPs.  Figure~\ref{zoom}a and b contain 
a small part of the full FOV shown with these two different 
contrasts. (A box outlines the subfield in Fig.~\ref{fullfov}.) 

The disk center was not fully devoid of magnetic activity
during the observation. There was a small nearby active 
region (AR NOAO 10971\footnote{It never developed fully, 
producing only a few pores during its peak activity.}), 
that the telescope pointing avoided. We discard the influence
of this AR on the observed region with the following arguments:
(1) a simultaneous {\it Hinode} SOT/NB Na{\sc i}\,5896\,\AA\
magnetogram including the outskirts of the AR and the SST FOV 
shows that the two of them  are well separated.
(2) The average unsigned magnetic flux density in our FOV (29\,G) 
is within the low range of values of the signals in a magnetogram 
of the quiet Sun taken under the same 
conditions ten days later, when no AR was 
present. This reference magnetogram yields a signal of 34$\pm$5\,G,
with the error bar representing the standard deviation among
the signals in randomly chosen subfields with the size of 
the SST FOV.

\section{Results}\label{results}

Figure~\ref{fullfov} shows the result of our selection.
The original G-band image has been enhanced in those
segmented structures identified as BPs. One can recognize 
super-granulation and meso-granulation scales,
although BPs are found throughout the FOV. We identify 
2380 BPs, which correspond to 0.97 BPs~Mm$^{-2}$. 
In the segmentation employed for BP selection, 
these BPs  cover 0.89\,\% of the surface 
(the so-called filling factor $f$). However, the filling
factor is an uncertain parameter. 
(The number density of BPs is far more robust.)
If the segmentation algorithm is used dilating the 
kernel by one pixel (roughly speaking, increasing 
the size of the individual BPs by 1 pixel),  
then the surface coverage becomes 2.2\,\%. This
other segmentation is as good as the original
one as judged from visual inspection.
The segmentation algorithm underestimates the area covered by the 
large BPs. They are split into several individual objects, 
and the empty space left between them does not 
contribute to $f$ 
(compare the bright continuous filaments in 
Fig.~\ref{zoom}b with the corresponding broken 
segmentation in Fig.~\ref{zoom}c).
The splitting of one large object into several 
small pieces also enlarges our BP number density. 
However, this bias does not have a big
impact because large obsjects are 
mostly in the network,  and when network and IN are 
treated separately, the resulting number densities 
turn out to be very similar (see below). 

We have tried to separate the contribution of network 
and IN. This separation is based on the 
Ca~H images taken simultaneously with the G-band images.
They are smoothed with a 4\,\arcsec\ kernel, with 
the bright patches above a threshold representing the network.
The threshold was tuned to get a 10\,\% area coverage,
which is the kind of network surface coverage during solar minimum 
\citep[e.g.,][]{fou01,san04}. 
Number densities and filling factors
were separately re-computed for the pixels 
inside and outside the network thus defined. The number density of 
BPs and the filling factor of the IN are slightly smaller than the ones 
for the full FOV, but not so much. One gets 0.85~BPs~Mm$^{-2}$ and 
$f=0.77$\,\% when the full FOV rendered  0.97~BPs~Mm$^{-2}$ and 0.89\,\%, respectively.
This small difference is due to the fact that BPs are found throughout, 
rather than being concentrated in the network. 
The density in the network patches is 2.2~BPs~Mm$^{-2}$,
whereas $f=2.2$\,\%.

As we explain in \S~\ref{intro}, our work was originally
motivated to calibrate the automatic procedure by 
\citet{bov01,bov07}. A detailled comparison goes beyond the 
scope of this Letter, but the algorithm systematically 
underestimates the number of BPs, so that it yields 
0.18~BPs~Mm$^{-2}$. 
The patches associated to these BPs are substantially larger than those 
inferred here, so that $f$ remains $\simeq2$\,\%.

\section{Discussion}

According to the current paradigm, the BPs trace intense magnetic 
concentrations so, our observation constrains the fraction of 
quiet Sun plasma having kGs. Our measurement should be regarded as 
a lower limit since we only include secure BPs  (\S~\ref{description}), 
the area of large BPs is underestimated 
(\S~\ref{results}),  and 
diverse arguments suggest 
that some kGs are not bright \citep[e.g.,][]{san01,vog05,bec07}.
Moreover, we miss BPs since the detectability of the small
ones critically depends on the resolution \citep[e.g.][]{tit96}, 
and quiet Sun BPs are often at the resolution of the observation 
\citep[e.g.,][]{san04a,bov08}. In fact,
differences of angular resolution partly explain the lower 
number density of BPs previously reported in the literature 
(\S~\ref{intro}). Another part of the difference is due to the use of 
automatic algorithms based on 
single frames to detect the BPs,  which are still unable to 
identify the smallest and dimmest BPs in the images 
(\S~\ref{results}).

Even a 1\,\% filling factor in the form of kG fields represents
a big challenge. They are not yet produced by 
the current turbulent solar dynamo simulations \citep[][]{vog07},
and they are overlooked in recent studies of the quiet Sun
magnetic field strength distribution based on Stokes
spectro-polarimetry \citep[e.g.,][]{oro07,ase09}. 
Assuming that typical BPs have 1.5~kG field strengths, 
they contribute to the solar magnetic fields with an unsigned 
flux density between 13~G and 33~G depending on whether we 
adopt 0.9\,\% or 2.2\,\% for their filling factors. 
However,  \citet{oro07} report only 9.5~G,  with field strengths 
dominated by hG rather than kG fields.
The solution to this double puzzle may also have to do
with insufficient spatial resolution. On the
one hand, the amount of kG fields created by 
magneto-convection simulations 
increases with increasing spatial resolution  
\cite[e.g.,][]{bus08}. On the other hand,
the standard procedure of interpreting Stokes 
profiles does not acknowledge the existence of unresolved 
structuring. When it is acknowledged, then kG fields
with filling factors compatible with the 
limits posed here are found \citep[][]{vit10}. 

Figure~\ref{fullfov} shows BPs throughout. Their 
abundance can be emphasized by comparison
with the number of granules. There are some 0.82~granules~Mm$^{-2}$ 
\citep[e.g.,][]{mul00}, which implies that the quiet Sun has 
more BPs than granules, i.e., our image has 
1.2~ {\rm BPs~per~granule.} Yet another revealing 
reference to compare with is the largest sunspot-group 
ever photographed on the Sun. It seems to correspond to a 
complex active region observed on April 7, 1947, 
which occupied  $4.3\times 10^3$ millionths of solar 
hemisphere \citep{hog47}. This record-braking large
sunspot yielded only $f\simeq 0.2$\,\%, i.e.,  
one would need several ARs like this one to carry
the same unsigned flux as the quiet Sun BPs 
have at any time if they uniformly cover 
the solar surface.

%
%
\acknowledgements
The SST is operated in the Spanish Observatorio del Roque de los Muchachos  
by the Institute for Solar Physics of the Royal Swedish Academy of 
Sciences.
{\em Hinode} is a Japanese mission developed and launched 
by ISAS/JAXA, with NAOJ, NASA (US) and STFC (UK) as partners.
We thank A. de Vicente for his support on the 
Condor\footnote{http://www.cs.wisc.edu/condor/} 
workload management system
used during image restoration. 
The work has been partly funded by the Spanish Ministries 
of Education, and of Science and Innovation (projects 
AYA2007-66502, AYA2007-63881 and ESP2006-13030-C06-04),
and by the EC (SOLAIRE Network -- MTRN-CT-2006-035484).
{\it Facilities:} \facility{SST}, \facility{Hinode (SOT/NB)}

%

%

\end{document}